\documentclass[aps, pra, reprint, superscriptaddress]{revtex4-1}
\usepackage[utf8]{inputenc}
\usepackage{amsmath}
\usepackage{amssymb}
\usepackage{hyperref}
\usepackage{graphicx}
\usepackage{epstopdf}
\usepackage{dcolumn}
\usepackage{bm}
\usepackage{xspace}
\usepackage{verbatim}
\usepackage{color}

\DeclareMathOperator{\Tr}{Tr}

\hypersetup{colorlinks=true,linkcolor=blue,citecolor=blue, filecolor=blue,urlcolor=blue,breaklinks=true}

\begin{document}
\title{Mechanical oscillator thermometry in the nonlinear optomechanical regime}
%\red{Quantum nanomechanical oscillator thermometry assisted by spins and optical probes}}

\author{V. Montenegro}
\email{vmontenegro@uestc.edu.cn}
\affiliation{Institute of Fundamental and Frontier Sciences, University of Electronic Science and Technology of China, Chengdu 610051, China}

\author{M. G. Genoni}
\email{marco.genoni@fisica.unimi.it}
\affiliation{Quantum Technology Lab $\&$ Applied Quantum Mechanics Group, Dipartimento di Fisica {\em Aldo Pontremoli}, Universit\`a degli Studi di Milano, I-20133 Milano, Italia}

\author{A. Bayat}
\email{abolfazl.bayat@uestc.edu.cn}
\affiliation{Institute of Fundamental and Frontier Sciences, University of Electronic Science and Technology of China, Chengdu 610051, China}

\author{M. G. A. Paris}
\email{matteo.paris@fisica.unimi.it}
\affiliation{Quantum Technology Lab $\&$ Applied Quantum Mechanics Group, Dipartimento di Fisica {\em Aldo Pontremoli}, Universit\`a degli Studi di Milano, I-20133 Milano, Italia}

\date{\today}
\begin{abstract}
Optomechanical systems are promising platforms for controlled light-matter interactions. They are capable of providing several fundamental and practical novel features when the mechanical oscillator is cooled down to nearly reach its ground state. In this framework, measuring the effective temperature of the oscillator is perhaps the most relevant step in the characterization of those systems. In conventional schemes, the cavity is driven strongly, and the overall system is well-described by a linear (Gaussian preserving) Hamiltonian. Here, we depart from this regime by considering an undriven optomechanical system via non-Gaussian radiation-pressure interaction. To measure the temperature of the mechanical oscillator, initially in a thermal state, we use light as a probe to coherently interact with it and create an entangled state. We show that the optical probe gets a nonlinear phase, resulting from the non-Gaussian interaction, and undergoes an incoherent phase diffusion process. To efficiently infer the temperature from the entangled light-matter state, we propose using a nonlinear Kerr medium before a homodyne detector. Remarkably, placing the Kerr medium enhances the precision to nearly saturate the ultimate quantum bound given by the quantum Fisher information. Furthermore, it also simplifies the thermometry procedure as it makes the choice of the homodyne local phase independent of the temperature, which avoids the need for adaptive sensing protocols.
\end{abstract}
\maketitle
\section{Introduction}
%Optomechanical:
Optomechanical systems have emerged as a formidable platform for the control and manipulation of light-matter interactions in quantum technologies~\cite{1.general-review-optomechanics-1,BowenMilburn}. From a fundamental perspective, they allow for preparing a superposition of quantum states of a macroscopic object~\cite{2.macro-superposition-1, 2.macro-superposition-2}, production of non-classical states for the light~\cite{3.optical-nc-1, 3.optical-nc-2} and the mechanics~\cite{4.mechanics-nc-1}, and may lead even to the detection of the quantum nature of gravity~\cite{5.quantum-gravity-1, 5.quantum-gravity-2}. Practically, the optomechanical systems can render hybrid architectures for quantum networking schemes~\cite{6.quantum-network-1}, the possibility of quantum state transfer~\cite{7.quantum-state-transfer-1} and quantum distillation~\cite{8.distillation-1}, and serve as a sensor for detecting small forces~\cite{9.force-1}, displacements~\cite{10.displacement-1}, masses~\cite{11.mass-1}, and accelerations~\cite{12.accelerometer-1, 12.accelerometer-2} with unprecedented precision. A crucial necessity for most of the above schemes is to possess the mechanical oscillator near its ground state~\cite{1.general-review-optomechanics-1, 13.importance-1}. Typically, the mechanical part operates at frequencies ranging from $1$ MHz to $1$ GHz~\cite{1.general-review-optomechanics-1}. This means that sophisticated cooling techniques are inevitable for reaching the mechanical ground state~\cite{14.ground-state-1, 14.ground-state-2, 14.ground-state-3, 14.ground-state-4}. To certify the success of any cooling procedure, it is of paramount importance to measure the temperature of the system precisely.

\begin{figure}[t]
\includegraphics[width=0.85 \linewidth]{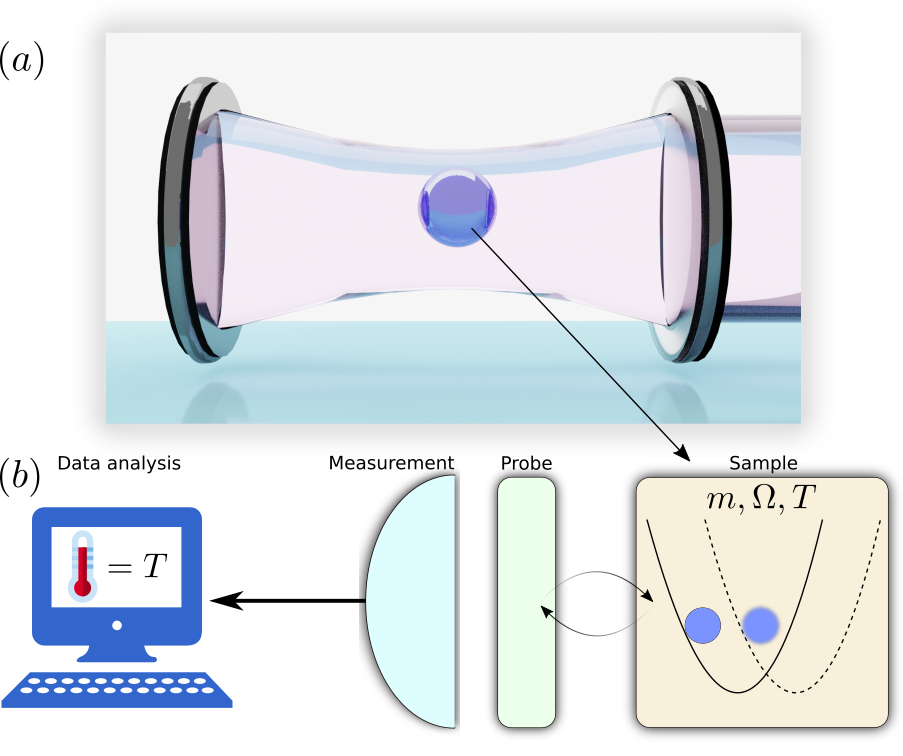}
\caption{(a) Schematic diagram of optomechanical system. (b) General procedure for the estimation of the oscillator's temperature $T$. The mechanical object (sample) of mass $m$, temperature $T$, and frequency $\Omega$ is probed by a coherent signal, interacting nonlinearly with the oscillator. To infer $T$, we suggest a feasible measurement 
scheme based on homodyne detection, which delivers nearly optimal thermometry performances.}\label{fig:model}
\end{figure}

%Thermometry:
Thermodynamical quantities (including temperature) are challenging to define, measure, and manipulate at quantum level~\cite{15.quantum-thermo-1}, which may even lead to reformulating the laws of thermodynamics~\cite{16.new-laws-thermo-1,16.new-laws-thermo-2,16.new-laws-thermo-3,16.new-laws-thermo-4,17.anna-review}. Concerning temperature, two main approaches may be identified for thermometry in the quantum domain: (i) the search for the optimal observable to be measured on the sample to extract information about temperature, and (ii) the design and the optimization of a {\em probing} technique, 
where the sample is let to interact with an external probe, which is then measured to extract information about the temperature of the sample.
The first approach~\cite{18.first-approach1,18.first-approach2,18.first-approach3,Campbell2018} is the most natural procedure for estimating temperature and
the optimal observable turns out to be the energy, as it happens in classical physics. However, this approach may be very demanding, as it 
requires access to the entire system, measuring its energy and having full knowledge of the spectrum. In the second approach, a small quantum probe interacts with the system without causing much disturbance and is then measured. Here we may distinguish two main strategies: one may consider a probe that interacts with the system  for a long time to reach equilibrium. Measuring the probe will the provide information about the temperature of the system~\cite{19.second-approach-1, 19.second-approach-2, 19.second-approach-3}. However, satisfying these conditions for fragile quantum systems may not be an easy task in practice. Alternatively, 
one may consider a quantum probe interacting with the system for a limited time \cite{20.third-approach-1, 20.third-approach-2, 20.third-approach-3, 20.third-approach-4, Feyles2019, Gebbia2019, Mancino2020} and the temperature becomes 
encoded in the entangled non-equilibrium system-probe quantum state.
Even tracing out the system degrees of freedom, temperature information 
remains mapped onto the state of the probe and may be extracted using a suitable set of measurements. Interestingly, this non-equilibrium scenario may yield enhanced precision compared to the equilibrated probes~\cite{21.justification-1}. Indeed, for systems that are prone to decoherence, such as optomechanical systems, interrogating the probe on a short timescale seems to be the most suitable strategy for thermometry. 
Notice that for probes at equilibrium the measured quantity is the thermodynamical temperature of the sample {\em and the probe}, while for out-of-equilibrium probes one just estimates a parameter of the probe density matrix, which turns out to be determined by the initial temperature of the sample.

%Optomechanical Technique
Currently, the dominant scheme for thermometry in optomechanical systems is based on the measurement of the so-called motional sidebands asymmetry ratio, i.e., $\bar{n}/(\bar{n} + 1)$ (with $\bar{n}$ being the mean phonon number)~\cite{22.sideband-asymmetry-1, 22.sideband-asymmetry-2, 22.sideband-asymmetry-3, 23.ratio-1}. Since this technique involves a cavity being strongly driven, the optomechanical system is typically linearized, and thus, the intrinsic nonlinear nature of the radiation-pressure optomechanical interaction cannot be addressed. In addition, even though heavily used in experiments, the motional sideband asymmetry technique may not provide the ultimate precision for thermometry. Therefore, developing new techniques for measuring the temperature of a mechanical object at the quantum precision limit in the nonlinear regime, as quantified by the quantum Fisher information, is highly desirable.

%What do we do?
In this paper, we consider an optomechanical system where no driving field is present and operating in the nonlinear regime. Initially, the mechanical oscillator is at thermal equilibrium at an unknown temperature. By switching on the interaction between the mechanical oscillator and the probing light, temperature information may be mapped to the quantum state 
of light, and it may be extracted through optical measurements, see Fig.~\ref{fig:model}. We have three main results: (i) the temperature parameter is shown to be imprinted solely as a phase diffusion process in the optical state; (ii) the quantum precision limit, set by the quantum Fisher information, is nearly saturated by placing a nonlinear Kerr medium before a homodyne detector; and (iii) by properly choosing the Kerr nonlinearity, the measurement basis becomes independent of temperature, avoiding complex adaptive sensing protocols. Our protocol is distinct from previous proposals as it neither relies on Gaussian interactions nor needs for adjustments of detunings~\cite{24.previous-works-1, 24.previous-works-2, 24.previous-works-3, 24.previous-works-4, 24.previous-works-5}.        

%Organization of the article.
The rest of the article is organized as follows: In Sec.~\ref{sec:preliminaries}, we briefly introduce the theory of quantum parameter estimation, for which we stressed the main equations to be used in the single parameter estimation case. In Sec.~\ref{sec:themodel}, we derive the reduced density matrix of the light probe. Sec.~\ref{sec:qfi}, accounts for the study of the quantum Fisher information. In Sec.~\ref{sec:cfi}, we present the measurement strategy to be employed in order to achieve the ultimate quantum bound. Finally, we present the conclusions of our results in Sec.~\ref{sec:conclusions}. 

\section{elements of parameter estimation}\label{sec:preliminaries}
Quantum parameter estimation aims to determine one or multiple quantities of interest by performing appropriate measurements and exploiting and estimator algorithm. In this work, we focus on single parameter estimation, where the only quantity to estimate is the temperature $T$ of a mechanical oscillator, whereas the rest of the parameters are assumed to be known and fully controlled. The estimation procedure will ultimately infer the quantity of interest using two essential steps: (i) gathering data through performing a specific type of measurement; and (ii) feed the gathered data into an estimator to infer the value of the parameter. For any choice of a measurement basis, the precision of the estimation obeys the classical Cram\'{e}r-Rao inequality~\cite{25.classical-cramer-rao-1}
\begin{equation}
\mathrm{Var}[T] \geq \frac{1}{M \mathcal{F}_C(T)},\label{eq:classical_cramer}
\end{equation}
where $M$ is the total number of measurements, $\mathrm{Var}[T]$ is the variance of the estimated quantity, and $\mathcal{F}_C(T)$ is the so-called classical Fisher information obtained as~\cite{25.classical-cramer-rao-1, 26.quantum-parameter-estimation-1}
\begin{equation}
\mathcal{F}_C(T) = \int dx \frac{1}{p(x|T)} \left[\partial_T p(x|T) \right]^2. \label{eq:classical_fisher}
\end{equation}
In the above expression, $\partial_T := \partial/\partial_T$, and $p(x|T)$ is the conditional probability for a measurement outcome $x$ given the temperature $T$. The equality in Eq.~\eqref{eq:classical_cramer} can be achieved when the estimator is optimal. In the asymptotical regime, where the data set is large, it is proven that Bayesian algorithm provides the best estimator~\cite{27.estimation, 26.quantum-parameter-estimation-1}. One can further generalize the above classical inequality by optimizing upon all the possible Positive-Operator Valued Measure (POVM) $\{\Pi_x\}$ operators, where $\int dx \Pi_x = \mathbb{I}$. This extra optimization tightens the above bound and leads to the quantum Cram\'{e}r-Rao inequality~\cite{26.quantum-parameter-estimation-1}
\begin{equation}
\mathrm{Var}[T] \geq \frac{1}{M \mathcal{F}_Q(T)},\label{eq:classical_quantum}
\end{equation}
where 
\begin{equation}
\mathcal{F}_Q(T) := \mathrm{Tr}[\left(\partial_T\rho_T\right) L_T] = \mathrm{Tr}[\rho_T L_T^2] \geq \mathcal{F}_C,\label{eq:QFI-def}
\end{equation}
is the quantum Fisher information $\mathcal{F}_Q(T)$, $\rho_T$ is the density matrix parametrized on the oscillator's temperature $T$, and $L_T$ is the so-called Symmetric Logarithmic Derivative (SLD). By expressing the density matrix $\rho_T$ in spectral decomposition, one can provide an explicit form of the SLD as follows~\cite{26.quantum-parameter-estimation-1}:
\begin{equation}
L_T = 2 \sum_{n,m} \frac{\langle \psi_m | \partial_T \rho_T| \psi_n \rangle}{\varrho_m + \varrho_n} | \psi_m \rangle \langle \psi_n |,
\label{eq:SLD}
\end{equation}
where $\rho_T = \sum_n \varrho_n |\psi_n\rangle \langle \psi_n|$, and $\varrho_m + \varrho_n \neq 0$. With the above definition in Eqs.~\eqref{eq:QFI-def}-\eqref{eq:SLD}, it is straighforward to finally reach the quantum Fisher information on this particular basis
\begin{equation}
\mathcal{F}_Q = 2 \sum_{n,m} \frac{|\langle \psi_m | \partial_T \rho_T| \psi_n \rangle|^2}{\varrho_m + \varrho_n}.
\label{eq:qfi_pure}
\end{equation}
This is the definition which is employed throughout our numerical simulations.

\section{The model}\label{sec:themodel}
The standard nonlinear optomechanical Hamiltonian in the absence of external driving is ($\hbar = 1$):
\begin{equation}
\hat{H} = \Omega \hat{b}^\dagger \hat{b} - g_0 \hat{a}^\dagger\hat{a}(\hat{b}^\dagger + \hat{b}),\label{eq:opto-hamiltonian}
\end{equation} 
where we have switched to an appropriate frame rotating at the frequency of the optical mode $\hat{a}$. The mechanical oscillator of frequency $\Omega$ and mode $\hat{b}$ couples to the light field with strenght $g_0$ (see Refs.~\cite{12.accelerometer-2} for a brief review on some explicit expressions for $g_0$ as different physical setups are considered). Under this specific type of interaction, the mechanical oscillator's potential shifts its equilibrium position conditioned upon the eigenenergies $n$ of the number operator $\hat{a}^\dagger\hat{a}$~\cite{1.general-review-optomechanics-1, 3.optical-nc-1, 3.optical-nc-2}.

The mechanical oscillator is assumed to be initially in a mixed thermal state at temperature $T$, the parameter to be estimated. It is convenient to represent the oscillator state in coherent basis
\begin{equation}
\rho_\mathrm{M}(0) = \frac{1}{\pi \bar{n}} \int |\beta\rangle\langle \beta| e^{-\frac{|\beta|^2}{\bar{n}}}d^2\beta,
\end{equation}
where 
\begin{equation}
\bar{n} = \left(\mathrm{exp}\left[\frac{\Omega}{k_B T}\right] - 1\right)^{-1}
\end{equation}
is the phonon occupancy number and $k_B$ is the Boltzmann constant. Since $\bar{n}$ is an injective function of $T$, we will refer to the oscillator's temperature estimation either using $\bar{n}(T):=\bar{n}$ or $T$ indistinctibly.

Assuming full control of the light probe, we consider an initial pure state spanned in Fock basis with known coefficients $c_k \in \mathbb{C}$ as
\begin{equation}
\rho_\mathrm{L}(0) = \sum_{ n,m = 0}^\infty c_n c_m^* |n\rangle\langle m|.
\end{equation}
Therefore, the initial state of the system becomes
\begin{equation}
\rho(0) = \rho_\mathrm{L}(0)\otimes\rho_\mathrm{M}(0)
\end{equation}
The system undergoes a time evolution as
\begin{equation}
\rho(t) = \hat{U}(t) \rho(0) \hat{U}^\dagger(t).
\end{equation}
where the time evolution operator $\hat{U}(\tau)=\exp(- i \hat{H} \tau)$ has been found to be~\cite{3.optical-nc-1, 3.optical-nc-2}
%Contrary to conventional temperature estimation in classical physics, where one reads the probe once the probe-sample system have reached thermal equilibrium, here we are interested in exploiting the underlying quantum system dynamics before any thermalization takes place. Accessing the light probe is obtained through the reduced density matrix of such state as $\rho(t)_\mathrm{light} = \mathrm{Tr}[\hat{U}(t) \rho(t_0) \hat{U}^\dagger(t)]$ ---where the trace is performed over the mechanical oscillator's degrees of freedom. The 
\begin{equation}
\hat{U}(\tau) = e^{i(g \hat{a}^\dagger \hat{a})^2(\tau  - \sin \tau)} e^{ g \hat{a}^\dagger \hat{a} (\eta \hat{b}^\dagger - \eta^* \hat{b})} e^{i\tau\hat{b}^\dagger \hat{b}}.\label{eq:unitary-operator}
\end{equation}
In the formula above we rescaled the relevant Hamiltonian shown in Eq.~\eqref{eq:opto-hamiltonian} by the mechanical frequency $\Omega$, and consequently, we have defined $g := g_0/\Omega$, $\eta := 1 - e^{-i\tau}$, and $\tau := \Omega t$. Notice that the second exponential in the time evolution operator is a displacement operator acting on the mechanical subsystem conditioned upon the observable $\hat{a}^\dagger\hat{a}$, whereas the first and third exponentials are a nonlinear function of the photon number operator $\hat{a}^\dagger\hat{a}$ and a phase shift operating solely on the optical and the mechanical modes, respectively.
One can find that the bipartite density matrix as
\begin{multline}
\rho(\tau) = \sum_{ n,m = 0}^\infty c_n c_m^* e^{i g^2(n^2 - m^2)(\tau - \sin\tau)}|n\rangle \langle m|\otimes \\
\frac{1}{\pi \bar{n}}\int d^2\beta e^{-\frac{|\beta|^2}{\bar{n}}}e^{\frac{g(n - m)}{2}[\beta^* (e^{i\tau} - 1) - \beta(e^{-i\tau} - 1)]}  |\phi_n\rangle \langle \phi_m|,\label{eq:bipartite-dynamics}
\end{multline}
with coherent mechanical amplitude
\begin{equation}
|\phi_n\rangle := |\beta e^{-i\tau} + g n \eta \rangle.
\end{equation}
Finally, by performing the trace over the oscillator's degrees of freedom in~\eqref{eq:bipartite-dynamics}, one can obtain the following reduced density matrix for the light field
\begin{equation}
\rho_\mathrm{L}(\tau) = \sum_{\substack{n=0\\m=0}}^\infty c_n c_m^* \mathcal{C}_{n,m} |n\rangle \langle m|,\label{eq:state_optomechanics1}
\end{equation}
where
\begin{equation}
\mathcal{C}_{n,m} = e^{i g^2(n^2-m^2)(\tau - \sin\tau)} e^{g^2(m-n)^2(1+2\bar{n})(\cos\tau - 1)}.\label{eq:state_optomechanics2}
\end{equation}
The expressions in Eqs.~\eqref{eq:state_optomechanics1}-\eqref{eq:state_optomechanics2} are the main results of this section. As evident, there are two different components in $\mathcal{C}_{n,m}$. The first exponential term is a coherent phase arising from the non-Gaussian interaction and does not depend on the temperature. The second term, however, is a phase diffusion which depends on temperature. We will provide more discussions about the quantum state of the light probe in the next section.

\subsection{Features of the light probe}
It is worth noting that the parameter to be estimated, namely 
$\bar{n}$, only arises in one of the exponentials in the 
reduced density matrix of the light probe, given in Eq.~\eqref{eq:state_optomechanics1}. In turn, this exponential 
resembles the detrimental effect of phase diffusion.
The diffusion process may be of course 
described using Lindblad formalism~\cite{19.diffusion-1}, but 
it can also be expressed as resulting from the application of
a random phase shift $\hat{U}_\theta := e^{-i\theta\hat{a}^\dagger\hat{a}}$, with $\theta$ being a random number sampled from a Gaussian distribution with zero-mean and standard deviation $\Delta$ 
\cite{19.diffusion-2, 19.diffusion-3}: 
\begin{align}
\rho_{\mathrm{D}} & = \frac{1}{\sqrt{4\pi\Delta^2}} \int_\mathbb{R} d\theta  e^{-\frac{\theta^2}{4\Delta^2}}\hat{U}_\theta\, \rho \, \hat{U}_\theta^\dagger \notag \\
& = \sum_{n,m=0}^\infty c_nc_m^* e^{-2(n - m)^2\Delta^2}|n\rangle\langle m|.
\end{align}
This process results in a degrading of the off-diagonal terms in the eigenbasis of $\hat{a}^\dagger\hat{a}$, yet conserving the energy. Notably, it also mimics the more complex process arising from the full bipartite dynamics shown in Eq.~\eqref{eq:state_optomechanics1}, where the precise amount $g^2(n - m)^2(1+2\bar{n})(\cos\tau - 1)$ emerges as a consequence of the mechanical coherent overlapping $\langle \phi_m | \phi_n \rangle$ and the relative phase from the displacement operator $e^{g(n-m)[\beta^* (e^{i\tau} - 1) - \beta(e^{-i\tau} - 1)]/2}$.

\section{Quantum Fisher Information}\label{sec:qfi}
The quantum Fisher information $\mathcal{F}_Q$ in general is a function of the tunable parameters of the system. To achieve the best precision for temperature estimation, one has to: (i) maximize $\mathcal{F}_Q$ with respect to such tunable parameters; and (ii) find the optimal measurement basis to achieve the bound given by $\mathcal{F}_Q$. Since the eigenvalue problem for the reduced quantum state in Eq.~\eqref{eq:state_optomechanics1} is analytically intractable, we rely on numerical methods for computing the quantum Fisher information. Moreover, even though a general photon distribution $c_n$ was considered for the derivation of the light probe, throughout this work we focus on a readily accesible input light, namely a coherent state with amplitude $\alpha \in \mathbb{R}$, which results in
\begin{equation}
c_n = e^{-\frac{\alpha^2}{2}} \frac{\alpha^n}{\sqrt{n!}}.
\end{equation}

%%% Figure #2
\begin{figure}[t]
\includegraphics[width=\linewidth]{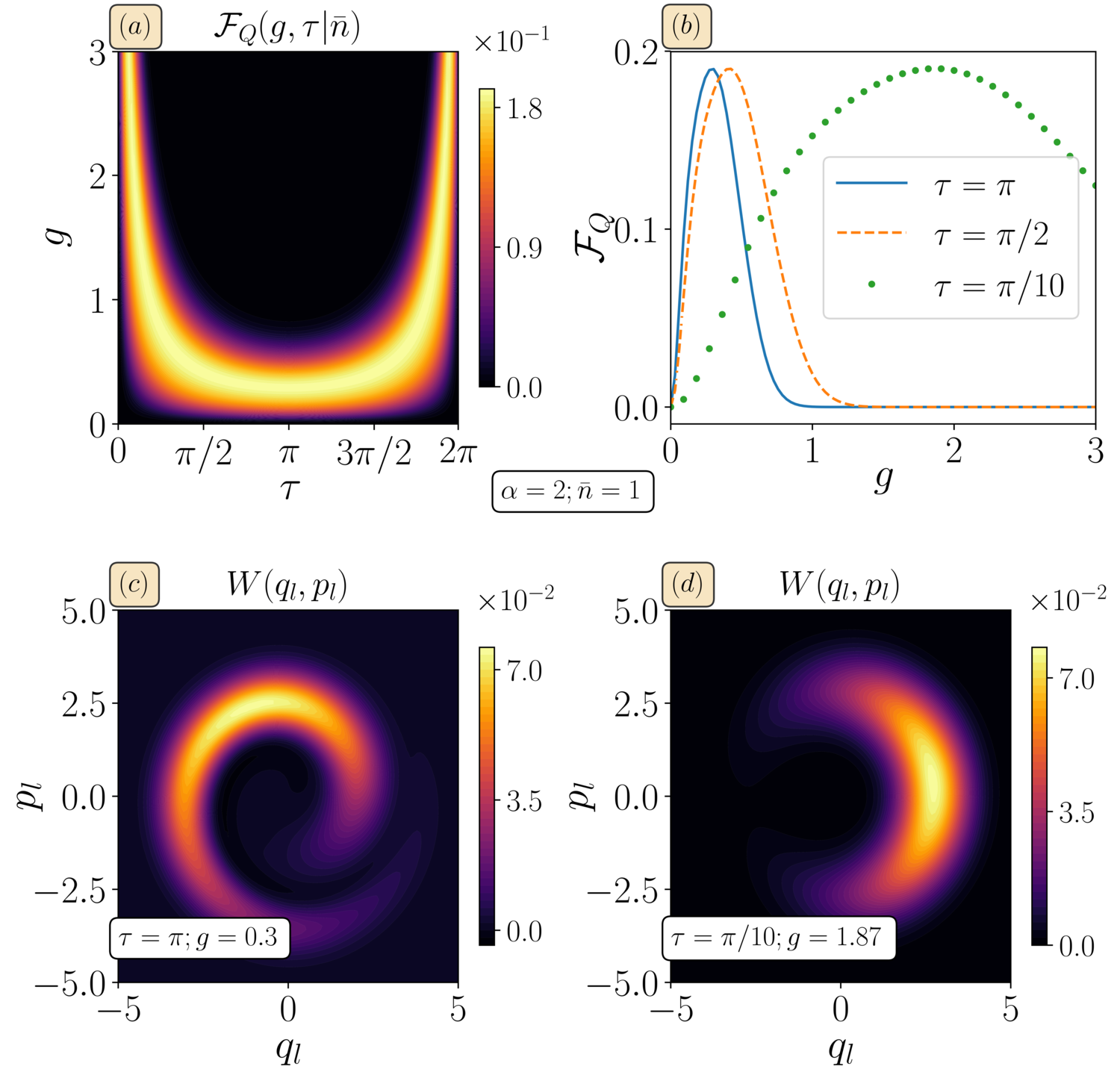}
\caption{(a) Quantum Fisher information $\mathcal{F}_Q(g,\tau|\bar{n})$ as functions of $g$ and $\tau$ for a given $\bar{n}$. As the figure shows, one can always adjust the set of parameters $g$ and $\tau$ in such a way that delivers maximal quantum Fisher information. In (b), we show the quantum Fisher information as function of $g$ for some interaction times $\tau$. As seen from the figure, an election of $\tau = \pi$ gives the lowest $g$ needed to reach maximal quantum Fisher information. Panels (c)-(d), show the Wigner function in phase space $\{q_l,p_l\}$ for the optical quantum state with the same maximal quantum Fisher information for times $\tau = \pi$ and $\tau=\pi/10$, respectively. An evident nonlinear phase as well as an incoherent phase diffusion is observed for $\tau = \pi$, whereas as the time decreases, say $\tau = \pi/10$, the nonlinear phase vanishes. Other values are $\alpha = 2$ and $\bar{n}=1$.}\label{fig:quantum-info}
\end{figure}
In Fig.~\ref{fig:quantum-info}(a) we show the quantum Fisher information $\mathcal{F}_Q(g,\tau|\bar{n})$ as functions of the optomechanical coupling $g$ and interaction time $\tau$ given a temperature $\bar{n}$. Without loss of generality, we have fixed the coherent amplitude $\alpha = 2$, as well as the phonon occupancy number to be $\bar{n} = 1$. As evident from the figure, there is a vast domain where the set of controlled parameters $\{g, \tau\}$ can always be adjusted such that the quantum Fisher information is maximal. This could be understood in terms of the effective phase diffussion exponential in the reduced density matrix in Eq.~\eqref{eq:state_optomechanics1}. To see this, let us first consider the limit of $\tau \ll 1$, under this limit the quantum state can be approximated as
% leading to the approximated state [$\sin(\tau) \approx \tau; \cos(\tau) \approx 1 - \tau^2/2$]:
\begin{equation}
\rho_\mathrm{L}(\tau) \stackrel{\tau \ll 1}{\approx} \sum_{\substack{n=0\\m=0}}^\infty e^{-\alpha^2} \frac{\alpha^{n+m}}{\sqrt{n!m!}}e^{-\frac{(g\tau)^2}{2}(m-n)^2(1+2\bar{n})} |n\rangle \langle m|,\label{eq:optics-approx}
\end{equation}
where the optomechanical coherent phase, arising from the non-Gaussian interaction, no longer plays a role and only the diffussion process is present. In this limit, as the dependende on $\{g, \tau\}$ is through their multiplication $g\tau$ by choosing a short interaction time $\tau$ a large $g$ is required for maximizing $\mathcal{F}_Q$. This is evident in the area in the $g$-$\tau$ plane for which the quantum Fisher information is maximal as shown in Fig.~\ref{fig:quantum-info}(a). On the other hand, as $\tau$ increases the relationship between $g$ and $\tau$ delivering maximal quantum Fisher information becomes more complex, this is because of the phase diffusion term $\mathrm{exp}[g^2(m-n)^2(1+2\bar{n})(\cos\tau - 1)]$. It follows that, for very small values of $g$, this term goes to one and dependence of $\bar{n}$ is lost. On the contrary, if $g$ is large, then the exponential term becomes vanishingly small, again losing its dependence on $\bar{n}$. Only for some intermediate values of $g$ the quantum Fisher information is maximal which is evident in Fig.~\ref{fig:quantum-info}(a). Interestingly, for an interaction time of $\tau = \pi$, one can maximize the quantum Fisher information by tuning the optomechanical strenght $g$ to its lower value. This election of the controlled parameters $\{g, \tau\}$ is of singular interest, as optomechanical systems in the nonlinear regime currently operates under weak radiation-pressure interaction coupling. Without loss of generality, from now on we will fix $\tau = \pi$, and $g = g_\mathrm{max}$ will correspond to the optomechanical coupling that brings the quantum Fisher information to its maximal value. To support the above, in Fig.~\ref{fig:quantum-info}(b), we show the quantum Fisher information $\mathcal{F}_Q$ as function of $g$ for different times $\tau$. As shown in the figure, different values of $g$ and $\tau$ lead to the same maximum value of the quantum Fisher information, for which $\tau = \pi$, as stated before, is the one delivering the lowest optomechanical coupling strenght $g$. 
 
To illustrate the differences between optical states with same quantum Fisher information, yet tuned with different choices of $g$ and $\tau$, we plot in Figs.~\ref{fig:quantum-info}(c)-(d) the quasiprobability Wigner function $W(q_l,p_l)$ in the phase space $\{q_l,p_l\}$ with associated quadratures of the light field. The Wigner function is numerically evaluated according to~\cite{20.qutip-1, 20.qutip-2}:
\begin{eqnarray}
\nonumber W(q_l,p_l) &=& \frac{1}{\pi}\int_{-\infty}^\infty\langle q_l + x|\rho_\mathrm{L}(\tau) |q_l - x \rangle e^{-2ip_l x} dx,\\
\nonumber &=& \sum_{n,m=0}^\infty \frac{e^{-\alpha^2} \alpha^{n+m}}{n! m! \sqrt{2^{n+m} \pi^3}} e^{-q_l^2} \mathcal{C}_{n,m} \\
\nonumber &\times& \int_{-\infty}^\infty e^{- (2ip_lx + x^2)}\mathcal{H}_m(q_l-x)\mathcal{H}_n(q_l+x)dx,\\
\end{eqnarray}
where we have used
\begin{equation}
\langle n | x \rangle = \frac{e^{-\frac{x^2}{2}} \mathcal{H}_n(x)}{\sqrt{2^{n} n!} \pi^{1/4}},
\end{equation}
with $\mathcal{H}_n(x)$ being the Hermite polynomials of order $n$.

In Fig.~\ref{fig:quantum-info}(c), we show the Wigner function of the light state when the interaction time is $\tau = \pi$ and $g \approx 0.3$. The nonlinear features arising from the non-Gaussian optomechanical interaction are apparent. The moderate-to-strong value of $g$ makes this case practically relevant, e.g. see Refs.~\cite{14.ground-state-3, feasibility-2, feasibility-3, feasibility-4, feasibility-5, feasibility-6} for experimental values. However, in this case, the non-Gaussian features may be challenging to detect by accessible measurement shemes, such as homodyne detection \cite{ngh1,ngh2}. On the other hand, as shown in Fig.~\ref{fig:quantum-info}(d), by considering $\tau = \pi/10$ and $g \approx 1.87$, the light state exhibits only phase diffusion features which may be more easily detected via homodyne detection. However, this comes at the cost of larger values of $g$ which may be experimentally unfeasible. Therefore, it is highly desirable to find a measurement strategy which operates at the small $g$ and yet is able to deliver excellent estimation performance.

%%% Figure #3
\begin{figure}[t]
\includegraphics[width=\linewidth]{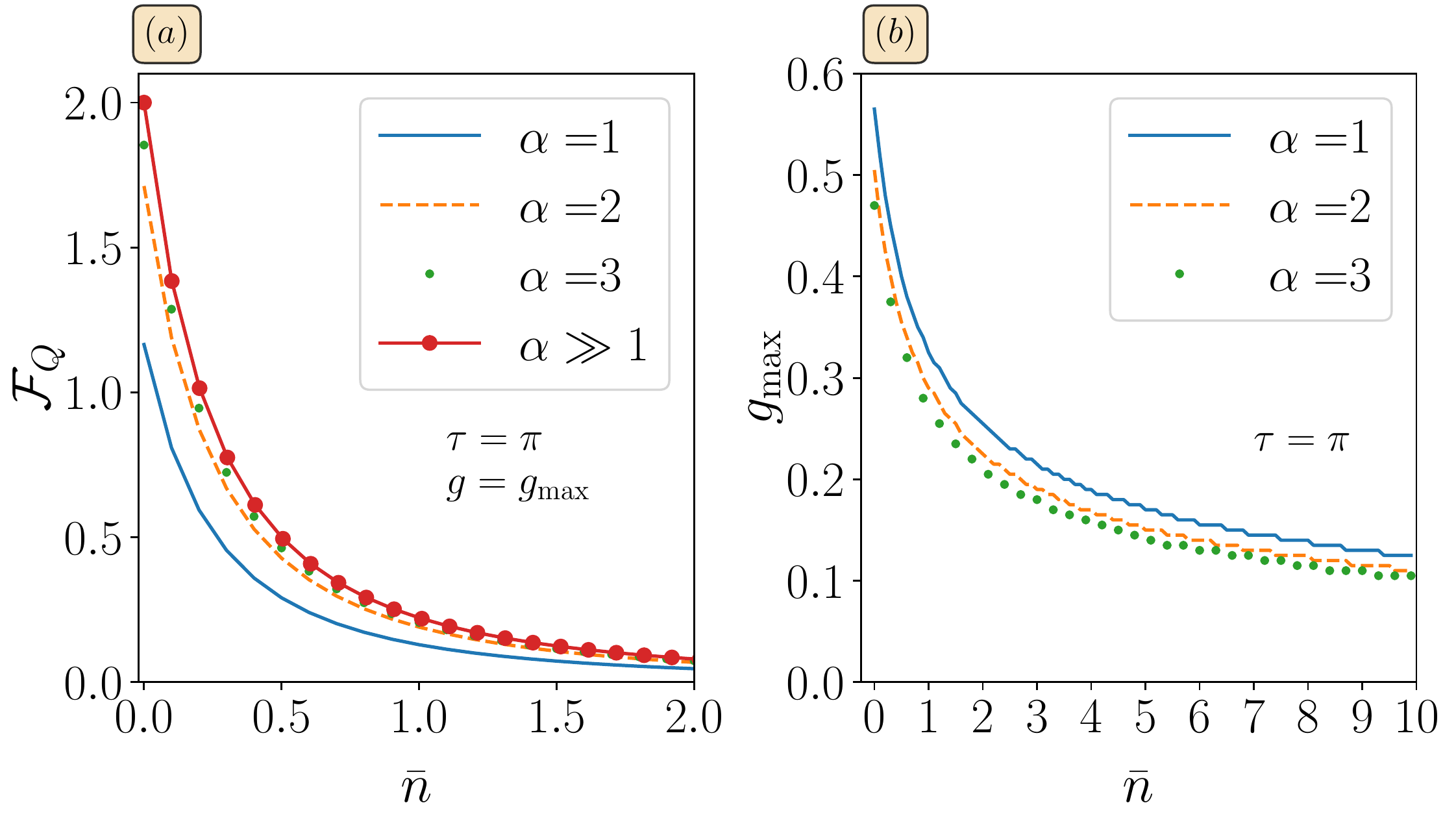}
\caption{(a) Quantum Fisher information as function of the oscillator's temperature $\bar{n}$ for different values of $\alpha$. The theoretical limit $\alpha \gg 1$, optimized for $\{g=g_\mathrm{max},\tau=\pi\}$, represents the maximum value at which the quantum Fisher information can reach for a given $\bar{n}$. Panel (b), shows the optomechanical coupling such that maximizes the quantum Fisher information $g_\mathrm{max}$ as function of $\bar{n}$.}\label{fig:qfi-nbar}
\end{figure}
In Fig.~\ref{fig:quantum-info} we kept $\bar{n}$ and $\alpha$ fixed. Now we investigate their impact on the quantum Fisher information. In Fig.~\ref{fig:qfi-nbar}(a), we show the quantum Fisher information as a function of the oscillator's temperature $\bar{n}$ for different values of the coherent amplitude $\alpha$. As the figure shows, the quantum Fisher information peaks at $\bar{n} = 0$ for any $\alpha$, while rapidly decreasing as the oscillator's temperature $\bar{n}$ grows. This can be intuitively understood as in the limit of high oscillator's temperature, i.e., $\bar{n} \gg 1$, the phase diffusion term $\mathrm{exp}[g^2(m-n)^2(1+2\bar{n})(\cos\tau - 1)]$, given in Eq.~\eqref{eq:state_optomechanics2}, goes to zero and weakly depends on the exact value $\bar{n}$, for all values of $\alpha$. In the opposite regime, i.e., $\bar{n} \ll 1$, the probe changes substantially as $\bar{n}$ varies. In other words, the variation of phonon excitations lead to a completely different optical phase diffusion term, and thus, one would expect better estimation and lower uncertainties for this quantity.

Furthermore, as the Fig.~\ref{fig:qfi-nbar}(a) shows, increasing the initial coherent amplitude $\alpha$ always benefits the precision in estimating the temperature of the oscillator, however, it quickly saturates for an initial number of photons above $\alpha^2 > 9$. In the limit of large $\alpha$, one can linearize the optomechanical Hamiltonian and the corresponding QFI can be analytically evaluated via the Gaussian formalism (see Appendix \ref{a:linearized} for more details about the derivation). By taking $\tau = \pi$, one gets
\begin{equation}
\mathcal{F}_Q \stackrel{\alpha \gg 1}{=} \frac{2}{(1 + 2\bar{n})^2}.\label{eq:limit-estimation}
\end{equation}
%In Appendix we provide a detailed discussion on the derivation of the above equation. 
Remarkably, as seen from the figure, even for  $\alpha^2 > 9$ one can almost achieve this limit.

As stated before, each point of the quantum Fisher information in Fig.~\ref{fig:qfi-nbar}(a) has been maximized using $\tau = \pi$ and $g = g_\mathrm{max}$. In Fig.~\ref{fig:qfi-nbar}(b), we depict the dependence of $g_\mathrm{max}$ as function of the temperature $\bar{n}$ for different coherent amplitudes $\alpha$. As the figure shows, comparable strong-to-moderate strenght of $g$ is observed for any chosen $\alpha$. The large values of $g$ when $\bar{n} \simeq 1$ can be intuitively explained as one requires stronger correlations between the light field and the oscillator in order to extract some information related to the mechanics.

\section{Classical Fisher Information}\label{sec:cfi}
The bound given by $\mathcal{F}_Q$ sets the ultimate precision limit allowed by quantum mechanics. Nonetheless, the quantum Cram\`er-Rao theorem does not explicitly provide the optimal measurement. In order to saturate the bound one needs to implement the optimal POVM, which is made by the set of projectors over the eigenstates of the SLD operator, i.e. $L_T$, in combination with optimal estimators. It is known that for large data sets a Bayesian estimator provides optimal estimation~\cite{27.estimation, 26.quantum-parameter-estimation-1}. One of the complex problems in quantum metrology is that the optimal measurement basis, computed from the eigenvetors of the SLD operator $L_T$, depend on the unknown parameter, here $\bar{n}$. The typical recipe for this problem is to follow complex adaptive approaches~\cite{21.adaptive-1, 21.adaptive-2, 21.adaptive-3, 21.adaptive-4, 21.adaptive-5, 21.adaptive-6} to update the measurement basis iteratively by extracting information about the exact value of the unknown parameter. In practice, to avoid such complexity, it is of significant importance if one can determine a fixed measurement basis which is independent of the unknown parameter and maximizes the quantum Fisher information. Therefore, in what follows we focus on determining an undemanding measurement which leads closely to the bound.
\subsection{Determining a feasible measurement}
%%% Figure #4
\begin{figure}[t]
\includegraphics[width=\linewidth]{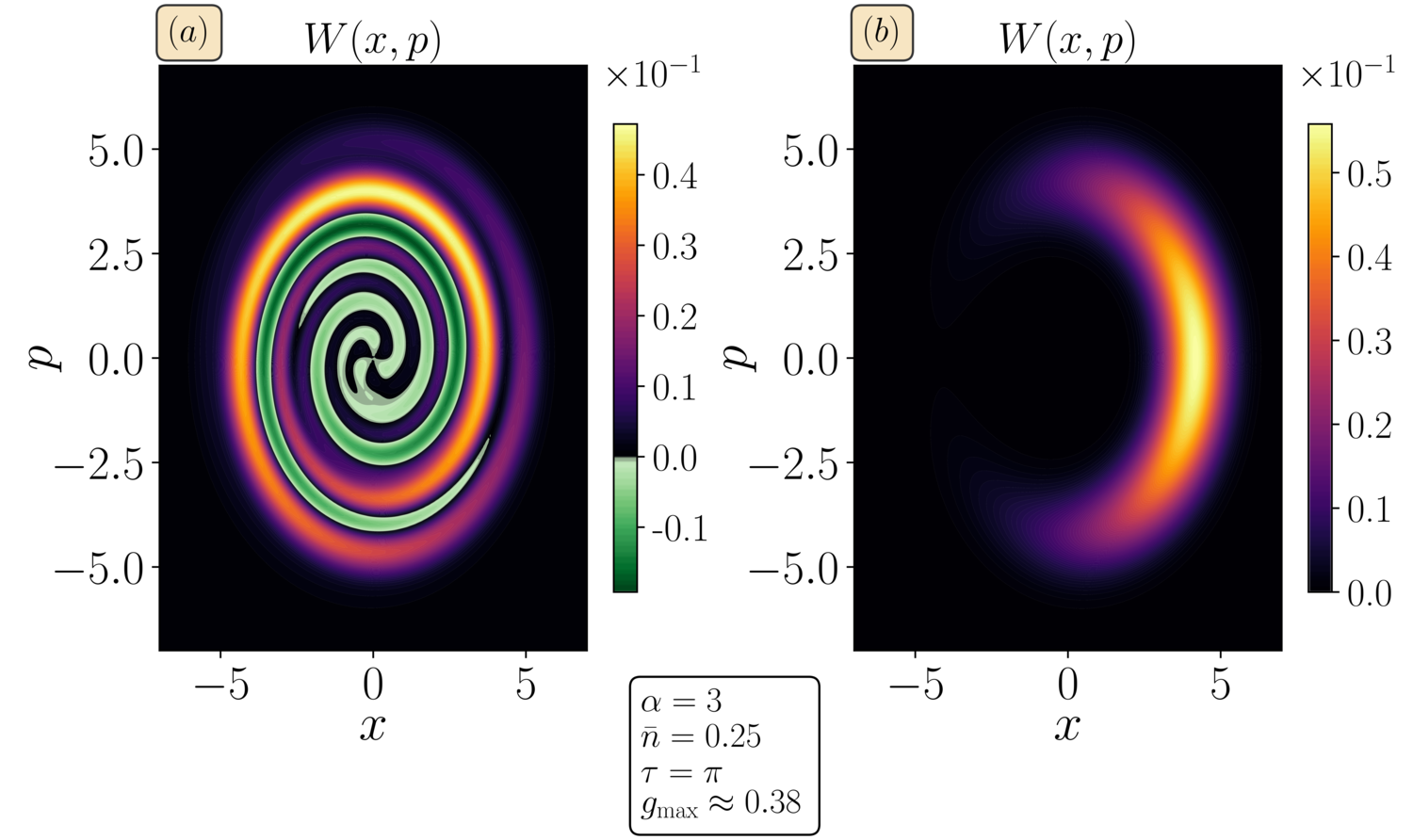}
\caption{(a) Wigner function of the light field for $\alpha = 3, \bar{n} = 0.25, \tau = \pi, g_\mathrm{max} \approx 0.38$. Significant negative values characterizes the nonclassical nature of the light field. (b) The optical state is led to interact with a Kerr medium of nonlinear strenght $\chi$. By a proper choice of $\chi = 2\pi g^2_\mathrm{max}$, one can fully suppress the intrinsic coherent nonlinear phase arising from the non-Gaussian optomechanical interaction.}\label{fig:undoing-kerr}
\end{figure}
As depicted in Fig.~\ref{fig:qfi-nbar}(a), the estimation of the oscillator's temperature delivers larger quantum Fisher information particularly for low phonon quanta excitations, say $0 \leq \bar{n} \leq 1$. Within this domain, the optical state may exhibit strong nonclassical features conditioned upon the coherent amplitude $\alpha$ and the strength of the optomechanical coupling $g$. For instance, in Fig.~\ref{fig:undoing-kerr}(a), we numerically evaluate the Wigner function of the light field for $\alpha = 3$ (near saturation of the quantum Fisher information shown in Eq.~\eqref{eq:limit-estimation}), and the achieved experimental mechanical oscillator ground state $\bar{n} = 0.25$~\cite{14.ground-state-1, 14.ground-state-2}. As the figure shows, the light field presents distinct nonclassical features, as evidenced by the ample negativity arising from the Wigner function. For this scenario, it is difficult to provide a true optimal measurement basis as the SLD may result in very complex measurement setups. Motivated by this, let us apply the following unitary operator on the quantum state of our probe
\begin{equation}
\hat{U}_\mathrm{K} = \mathrm{exp}\left[-\frac{i\chi}{2}(\hat{a}^\dagger\hat{a})^2\right],
\end{equation} 
where $\chi$ is a Kerr nonlinear tunable parameter. The reason behind the application of this nonlinear Kerr unitary operation is to modulate the temperature-independent phase in the quantum state of the probe, given in Eq.~\eqref{eq:state_optomechanics2}, to compensate the non-Gaussian effect of the Hamiltonian. The transformed state reads as
\begin{multline}
\tilde{\rho}_\mathrm{L}(\tau=\pi) = \hat{U}_\mathrm{K} \rho_\mathrm{L}(\tau = \pi)	 \hat{dU}_\mathrm{K}^\dagger \\
= e^{-\alpha^2}\sum_{\substack{n=0\\m=0}}^\infty \frac{\alpha^{n+m}}{\sqrt{n!m!}} e^{i (n^2-m^2)\left(\pi g_\mathrm{max}^2 - \frac{\chi}{2}\right)} \\
\times e^{-2g_\mathrm{max}^2(m-n)^2(1+2\bar{n})} |n\rangle \langle m|.
\end{multline}

The Wigner function of $\tilde{\rho}_\mathrm{L}(\tau=\pi)$ is depicted in Fig.~\ref{fig:undoing-kerr}(b) when $\chi$ is set to  $\chi = 2\pi g^2_\mathrm{max}$. Interestingly, by this choice the non-Gaussian phase is fully cancelled resulting in an entirely positive Wigner function. Indeed, it is the Wigner function of an initially Gaussian state subject ot phase diffusion.

To quantify the performance of this procedure, one has to evaluate the classical Fisher information $\mathcal{F}_C$ using homodyne detection preceded by a nonlinear Kerr medium, and compare it with the ultimate precision bound given by $\mathcal{F}_Q$. To evaluate the classical Fisher information $\mathcal{F}_C$ shown in Eq.~\eqref{eq:classical_fisher}, it is straightforward to obtain the conditional probability $p(x_{\Phi_\mathrm{LO}}|\bar{n})$ as 
\begin{multline}
p(x_{\Phi_\mathrm{LO}}|\bar{n}) = \mathrm{Tr}\left[ |x_{\Phi_\mathrm{LO}} \rangle \langle x_{\Phi_\mathrm{LO}}| \tilde{\rho}_\mathrm{L}(\tau=\pi) \right],\\
= \sum_{\substack{n=0\\m=0}}^\infty \frac{\alpha^{n+m}}{\sqrt{n!m!}} e^{i (n^2-m^2)\left(\pi g_\mathrm{max}^2 - \frac{\chi}{2}\right)}e^{-2g_\mathrm{max}^2(m-n)^2(1+2\bar{n})}\\
\times e^{-\alpha^2} e^{-x_{\Phi_\mathrm{LO}}^2} \frac{\mathcal{H}_m(x_{\Phi_\mathrm{LO}}) \mathcal{H}_n(x_{\Phi_\mathrm{LO}}) e^{i\Phi_\mathrm{LO}(m-n)}}{\sqrt{\pi 2^{(m+n)} m! n! }},
\end{multline}
where $|x_{\Phi_\mathrm{LO}}\rangle$ is the eigenvector of the rotated quadrature operator $\hat{x}_\phi$ with local oscillator phase $\phi$ defined as:
\begin{equation}
\hat{x}_{\Phi_\mathrm{LO}} = \frac{\hat{a} e^{-i\Phi_\mathrm{LO}} + \hat{a}^\dagger e^{i\Phi_\mathrm{LO}} }{\sqrt{2}}.
\end{equation}

%%% Figure #5
\begin{figure}[t]
\includegraphics[width=\linewidth]{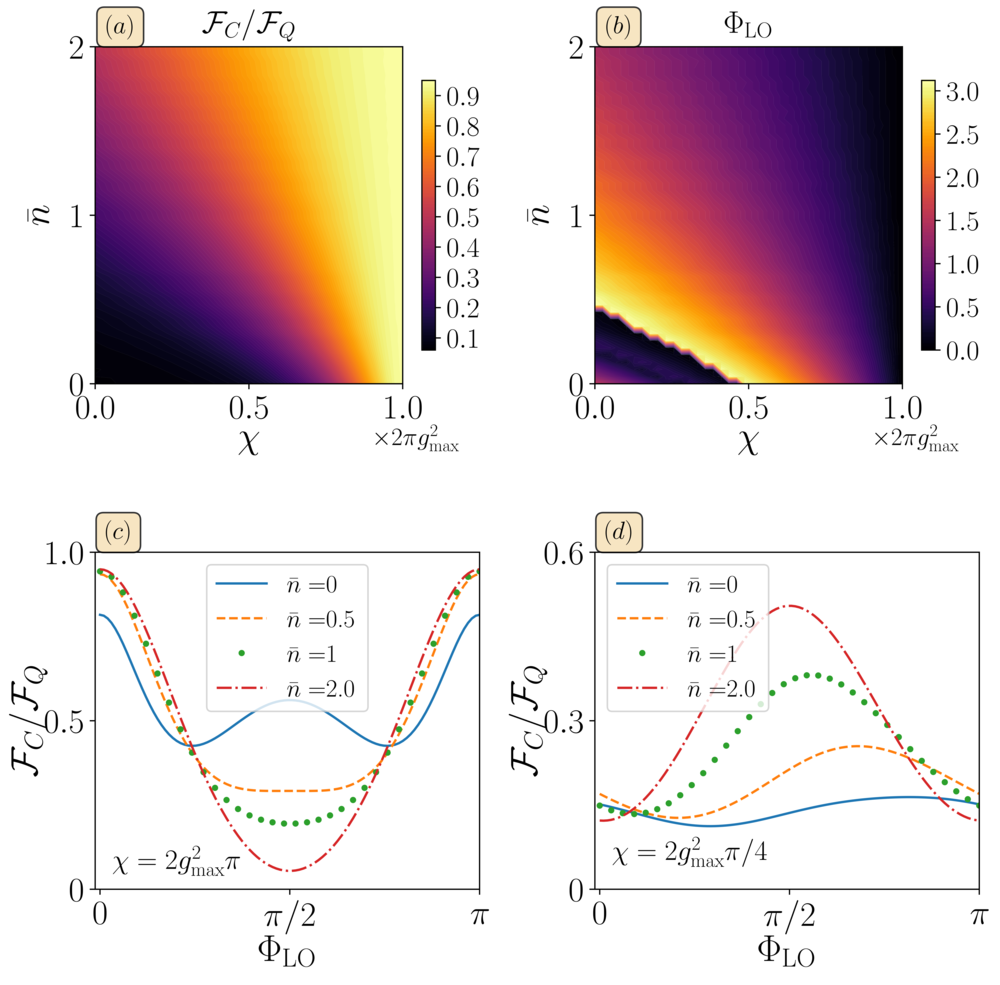}
\caption{(a) Fisher information ratio $\mathcal{F}_C/\mathcal{F}_Q$ as functions of the Kerr nonlinear strenght $0 \leq \chi \leq 2\pi g^2_\mathrm{max}$ and the oscillator's temperature $\bar{n}$. A proper tuning of $\chi$ and the known local oscillator phase $\Phi_\mathrm{LO}$ [see panel (b)] can lead to a Fisher information ratio up to $\mathcal{F}_C \approx 0.95 \mathcal{F}_Q$. (b) Local oscillator phase $\Phi_\mathrm{LO}$ as functions of the Kerr nonlinear strenght $0 \leq \chi \leq 2\pi g^2_\mathrm{max}$ and the oscillator's temperature $\bar{n}$. Notice that by tuning $\chi = 2g^2_\mathrm{max}\pi$ makes the measurement basis independent of the unknown parameter. In (c), we show the Fisher ratio $\mathcal{F}_C/\mathcal{F}_Q$ as a function of $\Phi_\mathrm{LO}$ for different values of $\bar{n}$ when the Kerr nonlinearity is tuned to $\chi = 2g_\mathrm{max}^2\pi$. As seen, the measurement basis becomes independent of the unknown parameter $\bar{n}$. Similarly, in panel (d), when the Kerr medium is tuned to $\chi = 2g_\mathrm{max}^2\pi/4$, the measurement basis depends on $\bar{n}$, as the peak of the Fisher ratio changes as the temperature varies.}\label{fig:cfi}
\end{figure}
In Fig.~\ref{fig:cfi}(a), we compute the Fisher information ratio $\mathcal{F}_C/\mathcal{F}_Q$ as a function of the Kerr nonlinear strength $\chi$ and the oscillator's temperature $\bar{n}$. Notice that the Kerr modulation ranges between $0 \leq \chi \leq 2\pi g^2_\mathrm{max}$, i.e., from no Kerr medium interaction to the value which cancels the phase from non-Gaussian interaction completely. As it is evident from the figure, the best performance is achieved when the $\chi = 2\pi g^2_\mathrm{max}$, for which $\mathcal{F}_C/\mathcal{F}_Q$ reaches a near-optimal ratio of $\sim 0.95$. 

Two relevant cases are pertinent to explore. On the one hand, for low phonon quanta excitations $\bar{n} \approx 0$, performing the homodyne detection step without any Kerr modulation $\chi = 0$ leads to a low Fisher information ratio about $\sim 0.1$ ---while letting the system to interact with a Kerr medium of strength $\chi = 2\pi g^2_\mathrm{max}$ one gains much information up to a Fisher ratio of $\sim 0.95$. This result can be understood as estimating such values of $\bar{n} \approx 0$ demands stronger optomechanical couplings, which then enables major nonclassical features arising from the non-Gaussian character of the Hamiltonian [see Fig.~\ref{fig:undoing-kerr}(a)]. Thus, to obtain better performances in the homodyne detection scheme, one requires to cancel the non-Gaussian phase contribution, in Eq.~\eqref{eq:state_optomechanics2}, significantly. On the other hand, for larger values of $\bar{n}$, i.e., $\bar{n} \geq 1$, even modest values of Kerr nonlinearity is enough to achieve large $\mathcal{F}_C/\mathcal{F}_Q$ ratio.

A crucial point in the above procedure for determining $\bar{n}$ is to fix $\Phi_\mathrm{LO}$, which specifies the homodyne measurement. If the optimized value of $\Phi_\mathrm{LO}$ depends on $\bar{n}$, which is unknown, then one has to resort in an adaptive approach. In such procedure, one has to acquire some prior information about $\bar{n}$ using non-optimal measurements, i.e., taking any value for $\Phi_\mathrm{LO}$, and then use the estimated value of $\bar{n}$ for updating the $\Phi_\mathrm{LO}$ for a better estimation in the next iteration. By repeating this for a few iterations, one can eventually tune $\Phi_\mathrm{LO}$ near its optimal value. It is highly desirable to find an optimal measurement independent of the parameter of interest, here $\bar{n}$. To investigate this, in Fig.~\ref{fig:cfi}(b), we plot the optimal $\Phi_\mathrm{LO}$ as a function of $\bar{n}$ and $\chi$. In general, for any choice of $\chi$, the optimal local phase $\Phi_\mathrm{LO}$ varies as $\bar{n}$ changes. Remarkably, by tuning $\chi = 2g_\mathrm{max}^2\pi$, which fully cancels the effect of the non-Gaussian optomechanical interaction, the optimal $\Phi_\mathrm{LO}$ becomes zero for any value of $\bar{n}$. This important observation shows that by using a Kerr nonlinear medium with $\chi = 2g_\mathrm{max}^2\pi$ one single measurement basis can detect $\bar{n}$ over a wide range of values, avoiding complex adaptive measurement methods. To show this more concretely, in Fig.~\ref{fig:cfi}(c), we plot the Fisher ratio $\mathcal{F}_C/\mathcal{F}_Q$ as a function of $\Phi_\mathrm{LO}$ for different values of $\bar{n}$ when the Kerr nonlinearity is tuned to $\chi = 2g_\mathrm{max}^2\pi$. As the figure shows, the maximum efficiency is achieved for $\Phi_\mathrm{LO} = 0$ or $\Phi_\mathrm{LO} = \pi$ for all values of $\bar{n}$. For the sake of completeness, in Fig.~\ref{fig:cfi}(d), we plot the $\mathcal{F}_C/\mathcal{F}_Q$ as a function of $\Phi_\mathrm{LO}$ when $\chi$ is tuned to a non-optimal value $\chi = 2g_\mathrm{max}^2\pi/4$ for various values of $\bar{n}$. As evident from the figure, for different values of $\bar{n}$ the peak of the curve varies, making an adaptive strategy essential.

It is also interesting to briefly discuss what happens in the limit of large $\alpha$ (i.e. for $\alpha \gg 1$). As we explain in Appendix~\ref{a:linearized}, if one considers the linearized optomechanical Hamiltonian, the classical Fisher information $\mathcal{F}_C$  for any Gaussian (general-dyne) measurement \cite{GenoniDiffusone,Serafozzi}, and thus comprising the special case of homodyne detection, can be analytically evaluated. Remarkably one shows that $\mathcal{F}_C$ goes to zero in the limit $\alpha \gg 1$ for any choice of the measurement. Non-Gaussian measurements are thus going to be necessary not only to attain the ultimate limit set by the QFI in Eq. (\ref{eq:limit-estimation}), but, in the limit of large $\alpha$, also to obtain a non-zero information about the temperature.

\section{Concluding remarks}\label{sec:conclusions}
In this paper, we have suggested a scheme for measuring the 
temperature of a mechanical oscillator, initially in a thermal state, 
using coherent light as a probe when the optomechanical system operates in the nonlinear regime. Remarkably, our scheme reaches precision, which almost saturates the quantum bound quantified by quantum Fisher information. To support our results, we analytically derive the temporal evolution of the reduced density matrix of the light probe, in which we find two different contributions: (i) a coherent phase due to the intrinsic non-Gaussian interaction term; and (ii) an incoherent diffusion process. Remarkably, the phase diffusion contribution is the only one encoding the mechanical oscillator's temperature. This suggests that the estimation performs better at low phonon quanta excitations, i.e., low temperature, as increasing the mean phonon number leads toward a complete loss of information regarding the optical phase. The key part of our protocol to achieve quantum-limited precision is to place a nonlinear Kerr medium before the homodyne detector. The introduction of this medium significantly increases the precision as it helps to cancel the temperature-independent coherent phase of the light probe. Hence, the measurement outcomes are solely determined by the incoherent diffusion process which encodes the initial temperature of the mechanical oscillator. Remarkably, by choosing the Kerr nonlinearity to fully cancel the coherent phase the local phase of the homodyne detection becomes independent of the temperature. This significantly simplifies the thermometry procedure, as it avoids the use of complex adaptive sensing methods.

\section*{Acknowledgements}
AB acknowledges the National Key R\&D Program of China, Grant No.
2018YFA0306703. VM thanks the Chinese Postdoctoral Science Fund for grant 2018M643435. MGG acknowledges support  from  a  Rita  Levi-Montalcini fellowship of MIUR. MGAP is member of INdAM-GNFM.

\appendix
\section{QFI for the linearized optomechanical Hamiltonian}\label{a:linearized}
In the case where the cavity field is prepared in a coherent state with large amplitude $\alpha \gg 1$, the optomechanical Hamiltonian in Eq. (\ref{eq:opto-hamiltonian}) can be linearized, i.e. it can be written as \cite{BowenMilburn}
\begin{align}
\hat{H}_{\sf lin} = \Omega \hat{b}^\dag \hat{b} -  g_0 \alpha (\hat{\tilde{a}} +\hat{\tilde{a}}^\dag) (\hat{b} + \hat{b}^\dag) \,,
\end{align}
where we have introduced the fluctuation of the cavity field operator around its mean value $\hat{\tilde{a}} = \hat{a} - \alpha$ (with $\alpha \in \mathbb{R}$) and we have neglected the nonlinear terms that in fact are not multiplied by $\alpha$.
By introducing the quadrature operators for the cavity field, $\hat{X} = (\hat{\tilde{a}}+ \hat{\tilde{a}}^\dag)/\sqrt{2}$, $\hat{Y} = i (\hat{\tilde{a}}^\dag - \hat{\tilde{a}})/\sqrt{2}$ and for the mechanical oscillator $\hat{Q} = (\hat{b}+ \hat{b}^\dag)/\sqrt{2}$, $\hat{P} = i (\hat{b}^\dag - \hat{b})/\sqrt{2}$, and by defining the vector of operators 
$\hat{\bf r} = ( \hat{X}, \hat{Y}, \hat{Q}, \hat{P})^{\sf T}$, one can rewrite the linearized Hamiltonian as 
\begin{align}
\hat{H}_{\sf lin} &= \frac{\Omega}{2} (\hat{Q}^2 + \hat{P}^2) -  2 g_0 \alpha \hat{Q}\hat{X} \,, \\
&=\frac{1}{2} \hat{\bf r}^{\sf T} H_{\sf lin} \hat{\bf r} \,,
\end{align}
where we have introduced the matrix 
\begin{align}
H_{\sf lin} = 
\left(
\begin{array}{c c c c}
0 & 0 & -2 g_0 \alpha & 0 \\
0 & 0 & 0 & 0 \\
- 2 g_0 \alpha & 0 & -\Omega & 0 \\
0 & 0 & 0 & \Omega
\end{array}
\right) \,.
\end{align}
As the Hamiltonian is quadratic in the bosonic operators, we can exploit the Gaussian formalism \cite{GenoniDiffusone,Serafozzi}: given that the initial state is a Gaussian state, one can describe the whole dynamics via the first moment vector and covariance matrix of the quantum state $\rho$, defined as
\begin{align}
\bar{\bf r} &= \Tr[ \rho \hat{\bf r} ] \, , \\
\boldsymbol{\sigma} &= \Tr[\rho \{ \hat{\bf r} - \bar{\bf r} , (\hat{\bf r} - \bar{\bf r})^{\sf T} \}] \,.
\end{align}
In our problem the initial state of the system is indeed Gaussian, being $\rho(0) = |\alpha\rangle\langle \alpha| \otimes \rho_M(0)$, corresponding to a zero first moment vector and a covariance matrix
\begin{align}
\boldsymbol{\sigma}(0) &= 
\left(
\begin{array}{c c c c}
1 & 0 & 0 & 0 \\
0 & 1 & 0 & 0 \\
0 & 0 & 2 \bar{n} +1 & 0 \\
0 & 0 & 0 & 2 \bar{n} + 1
\end{array}
\right) \,.
\end{align}
In the Gaussian formalism the unitary dynamics is described by a symplectic matrix that can be obtained via the formula $S(t) = \exp\{\omega H_{\sf lin} t\}$, where $\omega = \bigoplus_{j=1}^2 i \sigma_y$ denotes the symplectic form. In particular the covariance matrix (that encodes all the information about the temperature $\bar{n}$) evolves as $\boldsymbol{\sigma}(t) = S(t) \boldsymbol{\sigma}(0) S(t)^{\sf T}$. Performing the partial trace over the mechanical oscillator degrees of freedom, in the Gaussian formalism simply corresponds to take the $2 \times 2$ submatrix corresponding to the cavity field operator, that is
\begin{align}
\boldsymbol{\sigma}_L(\tau) = 
\left(
\begin{array}{c c}
1 & f(g, \alpha, \tau) \\
f(g, \alpha, \tau) & 1 + h(\bar{n},g, \alpha,\tau) + f(g, \alpha,\tau)^2 
\end{array}
\right) \nonumber
\end{align}
where we have introduced the functions
\begin{align}
f(g, \alpha, \tau) = 4 g^2 \alpha^2 ( \tau - \sin\tau ) \,, \\
h(\bar{n},g,\alpha, \tau) = 8 g^2 \alpha^2(1 - \cos\tau )(2 \bar{n} + 1)\,,
\end{align}
and we are considering the rescaled values $\tau = \Omega t$ and $g = g_0/\Omega$.
As mentioned above all the information about the temperature is encoded in the covariance matrix. As a consequence one can evaluate the corresponding QFI via the formula \cite{Pinel2013}
\begin{align}
\mathcal{F}_Q &= \frac{1}{2 (1 + \mu_L)} \Tr[\boldsymbol{\sigma}_L^{-1} ( \partial_{\bar{n}} \boldsymbol{\sigma}_L) \boldsymbol{\sigma}_L^{-1} (\partial_{\bar{n}} \boldsymbol{\sigma}_L) ]  \nonumber \\
&\,\,\,\, + 2 \frac{( \partial_{\bar{n}}  \mu_L )^2}{1 - \mu_L^4} \,,
\end{align}
where we have dropped the dependence on the evolution time $\tau$ and we have introduced the purity of the state $\mu_L=\Tr[\rho_L(\tau)^2] =  1/\sqrt{{\rm Det} \boldsymbol{\sigma}_L }$. By exploiting the formula for $\boldsymbol{\sigma}_L$, one obtains the analytical result
\begin{align}
\mathcal{F}_Q &= \frac{8 g^2 \alpha^2 (\cos \tau - 1)}{(2 \bar{n} + 1) \left[4 g^2 \alpha^2 (2 \bar{n}+1)( \cos \tau - 1) - 1 \right]} \,,
\end{align}
that, by fixing the evolution time $\tau=\pi$, reads
\begin{align}
\mathcal{F}_Q(\tau=\pi) &= \frac{16 g^2 \alpha^2 }{(2 \bar{n} + 1)(1 + 8 g^2 \alpha^2 (2 \bar{n}+1))} \,, \nonumber \\
&\stackrel{\alpha \gg 1}{=} \frac{2}{(1+2 \bar{n})^2 }\,.
\end{align}

It is also possible to evaluate the classical Fisher information $\mathcal{F}_C$ corresponding to any Gaussian ({\em general-dyne}) measurement performed on the cavity field. In fact any projective Gaussian measurement can be described itself by a (covariance) matrix \cite{GenoniDiffusone,Serafozzi}
\begin{align}
\boldsymbol{\sigma}_M = R(\theta) \left(
\begin{array}{c c}
z & 0 \\
0 & 1/z 
\end{array}
\right) R(\theta)^{\sf T} \,,
\end{align}
where 
\begin{align}
R(\theta) = \left(
\begin{array}{c c}
\cos\theta & \sin\theta \\
-\sin\theta & \cos\theta
\end{array}
\right) 
\end{align}
denotes a two-dimensional rotation matrix of angle $\theta$. In particular heterodyne detection, that is projection on coherent states, is obtained for $z=1$, while homodyne detection that is projection on the eigenstates of the quadrature $\hat{X}_\theta = \cos\theta \,\hat{X} + \sin\theta\,\hat{Y}$ is obtained by considering the limit $z\rightarrow 0$. The measurement outcome is in general represented by a two-dimensional vector ${\bf r}_m$, and only in the limit of homodyne detection (that is for $z\rightarrow 0$) corresponds effectively to a single-valued outcome. Its conditional probability distribution $p({\bf r}_m | \bar{n})$ is a Gaussian multi-variate probability distribution centred in the light first-moment vector ${\bf r}_L=\Tr[\rho (\hat{X}, \hat{Y})^{\sf T}]$ and with covariance matrix $\boldsymbol{\Sigma} = (\boldsymbol{\sigma}_L + \boldsymbol{\sigma}_M)/2$. As previously, only the covariance matrix depends on the parameter $\bar{n}$ and the corresponding classical Fisher information can be evaluated via the formula
\begin{align}
\mathcal{F}_C = \frac{1}{2} \Tr[\boldsymbol{\Sigma}^{-1} ( \partial_{\bar{n}} \boldsymbol{\Sigma}) \boldsymbol{\Sigma}^{-1} (\partial_{\bar{n}} \boldsymbol{\Sigma}) ] \,.
\end{align}
 Since the most general formula is too cumbersome, we report here only the result obtained by setting $\tau=\pi$ and by considering a generic homodyne detection of the quadrature $\hat{X}_{\theta}$, yielding
 \begin{widetext}
\begin{align}
\mathcal{F}_C = \frac{2 (4 g \alpha \sin\theta)^4}{\left[ \cos^2\theta - 4 \pi g^2 \alpha^2 \sin(2\theta) + \left( 1 + 16 g^2 \alpha^2 (1 + 2 \bar{n}) + 16 \pi^2 g^4 \alpha^4 \right) \sin^2\theta \right]^2 } \,.
\end{align}
\end{widetext}
%\begin{align}
%\mathcal{F}_C = \frac{2 (4 g \alpha \sin\theta)^4}{\left[ \cos^2\theta - f(g,\alpha,\pi) \sin(2\theta) + \left( 1 + h(\bar{n},g,\alpha,\pi) + f(g,\alpha,\pi)^2 \right) \sin^2\theta \right]^2 } \,.
%\end{align}
Both the above formula and the most general one go to zero in the limit $\alpha \gg 1$.
This clearly shows how in this regime any {\em Gaussian} measurement, including homodyne detection, will bring no information on the temperature, and thus one has to resort to non-Gaussian measurements such as the one based on a Kerr interaction, suggested in the main text.
%By studying in more detail the light field covariance matrix $\boldsymbol{\sigma}_L$, we notice that the only element that depends on the temperature $\bar{n}$ is the variance of any quadrature operator $\hat{Y}$ (that is the element $(2,2)$), via the function $h(\bar{n},g, \alpha, \tau)$. In the limit of $\alpha \gg 1$, one can neglect $h(\bar{n},g,\alpha, \tau)$ in the covariance matrix $\boldsymbol{\sigma}_L$ as $f(g, \alpha, \tau)^2 \gg h(\bar{n},g,\alpha, \tau)$ and thus $\boldsymbol{\sigma}_L$ becomes independent of $\bar{n}$. 

\end{document}